\documentclass[aps,prl,reprint,superscriptaddress]{revtex4-1}

\usepackage{hyperref}
\usepackage{graphicx}
\usepackage{enumerate}
\hypersetup{
    pdfnewwindow=true,      
    colorlinks=true,       
    linkcolor=blue,          
    citecolor=blue,        
    filecolor=blue,      
    urlcolor=blue           
}

\usepackage{subfigure}
\usepackage{amsmath}
\usepackage{amssymb}
\usepackage{amsfonts}

\usepackage{color}

\providecommand{\apj}[0]{ApJ}
\providecommand{\apjl}[0]{ApJ Lett.}
\providecommand{\apjs}[0]{ApJ Supp. Ser. }

\providecommand{\aap}[0]{A\&A}

\providecommand{\mnras}[0]{MNRAS}

\providecommand{\nat}[0]{Nature}
\providecommand{\prl}[0]{PRL}
\providecommand{\prd}{PRD}

\providecommand{\ssr}[0]{Space Sci. Rev.}

\def\A{A}
\def\nBG{\langle n^{\rm BG} \rangle}
\def\nBGmu{\langle n^{\rm BG} \rangle}

\def\nD{\langle n \rangle}
\def\nDmu{\langle n \rangle}

\def\Fn{\Psi}
\def\Fmu{\Psi}

\def\S{S}
\def\eff{w}
\def\en{\epsilon}
\def\Scut{S_{\rm cut}}
\def\Snet{S_{\rm net}}
\def\Eiso{E_{\rm iso}}
\def\Ediss{E_{\rm diss}}

\def\fad{f_{\rm ad}}
\def\tdec{t_{\rm dec}}
\def\tc{t_c}
\def\tauT{\tau_{\rm T}}
\def\sT{\sigma_{\rm T}}

\def\Eq{Equation}
\def\Eqs{Equations}

\def\beq{\begin{equation}}
\def\eeq{\end{equation}}

\newbox\grsign \setbox\grsign=\hbox{$>$} \newdimen\grdimen \grdimen=\ht\grsign
\newbox\simlessbox \newbox\simgreatbox \newbox\simpropbox
\setbox\simgreatbox=\hbox{\raise.5ex\hbox{$>$}\llap
     {\lower.5ex\hbox{$\sim$}}}\ht1=\grdimen\dp1=0pt
\setbox\simlessbox=\hbox{\raise.5ex\hbox{$<$}\llap
     {\lower.5ex\hbox{$\sim$}}}\ht2=\grdimen\dp2=0pt
\setbox\simpropbox=\hbox{\raise.5ex\hbox{$\propto$}\llap
     {\lower.5ex\hbox{$\sim$}}}\ht2=\grdimen\dp2=0pt
\def\simgt{\mathrel{\copy\simgreatbox}}

\begin{document}

\title{Detection Prospects for GeV Neutrinos from Collisionally Heated Gamma-ray Bursts with IceCube/DeepCore}

\author{I. Bartos}
\email{ibartos@phys.columbia.edu}
\thanks{Columbia Science Fellow}
\affiliation{Department of Physics, Columbia University, New York, NY 10027, USA}
\affiliation{Columbia Astrophysics Laboratory, Columbia University, New York, NY 10027, USA}
\author{A. M. Beloborodov}
\affiliation{Department of Physics, Columbia University, New York, NY 10027, USA}
\affiliation{Columbia Astrophysics Laboratory, Columbia University, New York, NY 10027, USA}
\author{K. Hurley}
\affiliation{University of California-Berkeley, Space Sciences Laboratory, Berkeley, CA 94720, USA}
\author{S. M\'arka}
\affiliation{Department of Physics, Columbia University, New York, NY 10027, USA}
\affiliation{Columbia Astrophysics Laboratory, Columbia University, New York, NY 10027, USA}

\begin{abstract}
Jet re-heating via nuclear collisions has recently been proposed as the main mechanism for gamma-ray burst (GRB) emission. Besides producing the observed gamma-rays, collisional heating must generate 10-100\,GeV neutrinos, implying a close relation between the neutrino and gamma-ray luminosities. We exploit this theoretical relation to make predictions for possible GRB detections by IceCube+DeepCore. To estimate the expected neutrino signal, we use the largest sample of bursts observed by BATSE in 1991-2000. GRB neutrinos could have been detected if IceCube+DeepCore operated at that time. Detection of 10-100~GeV neutrinos would have significant implications, shedding light on the composition of GRB jets and their Lorentz factors. This could be an important target in designing future upgrades of the IceCube+DeepCore observatory.
\end{abstract}


\maketitle

\noindent
{ \bf 1. Introduction} --- Cosmological gamma-ray bursts (GRBs) are expected to be efficient producers of neutrinos. There are at least three mechanisms for neutrino emission:

(1) The GRB central engine has a characteristic temperature comparable
to 10\,MeV and is expected to emit quasi-thermal ($\en\sim 30-50$\,MeV) neutrinos with luminosities $L_\nu\sim 10^{53}$\,erg\,s$^{-1}$
on a timescale of $1-10$\,s.
Similar neutrinos are generally produced by collapsing stars that form
neutron stars or black holes; they have been detected in SN~1987A~\cite{PhysRevLett.58.1490,PhysRevLett.58.1494}.
These relatively low-energy neutrinos can hardly
be detected from typical GRBs, because they occur at cosmological distances.

(2) GRB jets carry plasma with high Lorentz factors $\Gamma=100-1000$.
The jets are unsteady, and internal collisions between baryons
are expected to produce pions whose decay
leads to neutrino emission of energy $\en\sim \Gamma m_\pi c^2$ \cite{1994ApJ...427..708P,1999ApJ...521..640D,2000PhRvL..85.1362B,2000ApJ...541L...5M,2007A&A...471..395K}.
This energy falls in the $10-100$\,GeV range. Previous searches for multi-GeV neutrinos from GRBs did not have sufficient sensitivity. We show in this paper that the IceCube+DeepCore detector provides the required sensitivity.

(3) A fraction of ions in GRB jets may be accelerated to ultra-high
energies; in particular, Fermi acceleration in internal shocks was proposed
\cite{1995PhRvL..75..386W}. The existing upper limits \cite{2012Natur.484..351A} indicate that this mechanism is relatively
inefficient.

Here we discuss the prospects for detecting neutrinos produced by the
second mechanism --- by nuclear collisions in the GRB jet (see also \cite{2013arXiv1301.4236M,2012PhRvD..85j3009G}). Collisional heating was recently found to naturally produce the observed $\gamma$-ray spectra \cite{2010MNRAS.407.1033B,2011ApJ...738...77V} and can be the dominant radiative mechanism of GRBs. It implies a relation between the observed $\gamma$-ray emission and the expected 10-100\,GeV neutrino flux, and one can use
this relation to make predictions for possible
neutrino detections.


\vspace{4 mm}
\noindent
{ \bf 2. Collisional mechanism} --- 
The light curves of observed bursts suggest that the GRB jets are unsteady on timescales as short as 1\,ms, and their non-thermal spectra indicate that the
energy of internal bulk motions is dissipated and converted to radiation. This dissipation may occur above or below the jet photosphere. Observed spectra are in conflict with optically thin models (see e.g., \cite{2013ApJ...764..157B}); this suggests that the burst emission is produced mainly by dissipation
in the opaque, sub-photospheric region.

An efficient dissipative mechanism below the photosphere is provided by nuclear collisions
(e.g., \cite{1994ApJ...427..708P,1999ApJ...521..640D,2000PhRvL..85.1362B,2006MNRAS.369.1797R,2010MNRAS.407.1033B}).
  As long as  internal motions in the jet are
at least mildly relativistic, the collision energy $\epsilon_{\rm coll}$ is comparable to or exceeds the proton rest mass, $m_pc^2\approx 1$\,GeV. This energy is sufficient for pion production. If $\epsilon_{\rm coll}>1$\,GeV, mutiple pions are produced with comparable (mildly relativistic) momenta in their center-of-momentum frame. The pion decay generates energetic electrons, and the electrons radiate their energy via synchrotron emission and inverse Compton scattering, which involves a cascade of $e^\pm$ creation.
The cascade develops because the high-energy photons produced by inverse Compton scattering quickly collide with softer photons and
convert to $e^\pm$ pairs.
Using the known rates of collisional and radiative processes, one
can predict from first principles the
gamma-ray spectrum emerging at the jet photosphere. Detailed calculations of radiative transfer in the collisionally heated, expanding
jet gave GRB spectra consistent with observations \cite{2010MNRAS.407.1033B,2011ApJ...738...77V}. In particular, the position of the spectral peak, and the spectral slopes below and above the peak were found to agree with data.

In this model, the observed extended tail of the $\gamma$-ray spectrum is generated by pions produced in inelastic nuclear collisions. Pions quickly decay into particles of energy $\sim 10^2m_ec^2$; e.g. $\pi^+$ decay through reactions $\pi^+\rightarrow \mu^+ + \nu_\mu\rightarrow e^+ + \nu_e + \overline{\nu}_\mu + \nu_\mu$, and similar reactions describe the decay of $\pi^-$. Neutral pions decay into two high-energy photons. As a result, roughly half of the pion energy is emitted in neutrinos, and the other half is processed into radiation (and secondary $e^\pm$ pairs) through the cascade and synchrotron  emission.
  The radiative cooling of final products (electrons and positrons) occurs much faster than the expansion of the jet \cite{2010MNRAS.407.1033B}. Note also that the decay reactions are extremely fast, and radiative losses of intermediate particles (pions and muons) are negligible. Measured in the jet rest frame, the  lifetime of mildly relativistic $\pi^\pm$ is $\tdec\sim 3\times 10^{-8}$~s, and the lifetime of $\mu^\pm$ is $\tdec\sim 3\times 10^{-6}$~s. The cooling timescale $\tc$ of a particle of mass $m$ and elementary charge $e$ is $(m/m_e)^3$ longer than the electron cooling timescale, $t_{c,e}$.
   The typical value of $t_{c,e}$ in the collisionally heated region is
   $\sim 10^{-5}$~s
   \cite{2010MNRAS.407.1033B}, which implies $\tc\gg \tdec$ for pions and muons.

In each decay reaction, the energy is approximately evenly distributed between the decay products. Neutrinos therefore carry away a significant fraction $f_\nu\sim1/2$ of the energy dissipated in inelastic nuclear collisions (with $1-f_\nu$ given to radiation). Dissipation of energy $\Ediss$ 
in the opaque jet produces a GRB of energy
\beq
   E_\gamma=\fad\,(1-f_\nu) \Ediss,
\eeq
where $\fad<1$ describes the reduction in radiation energy due to adiabatic cooling in the expanding opaque jet below the photosphere. The adiabatic cooling factor for radiation produced at scattering (Thomson) optical depth $\tauT\gg 1$ and released at the photosphere is given by \cite{2011ApJ...737...68B}
\beq
   \fad(\tauT)=2\,\tauT^{-2/3}, \qquad \tau\gg 1.
\eeq
Note that the dissipated energy of internal bulk motions that is not converted to neutrinos tends to convert back into bulk kinetic energy via adiabatic cooling, leading to repeated dissipation.

The corresponding energy of the neutrino burst (which does not suffer any
adiabatic cooling) is
\beq
  E_\nu=f_\nu \Ediss.
\eeq
The ratio of neutrino and radiation burst energies (or their isotropic equivalents)
is given by
\beq
\label{eq:w}
   w=\frac{E_\nu}{E_\gamma}=\frac{f_\nu}{1-f_\nu}\,\frac{\overline{\tauT^{2/3}}}{2},
\eeq
where the line over $\tauT^{2/3}$ signifies the average over the region of
collisional dissipation.
Strongest collisional heating is expected at optical depths $\tauT\simgt \sigma_n/\sT\approx 20$ \cite{2010MNRAS.407.1033B}, where
$\sigma_n$ and $\sT$ are the nuclear and Thomson cross sections.
Therefore, the expected theoretical value for $w$ is 3 to 10.

The emitted neutrinos have energy comparable to $m_\pi c^2$ in the rest-frame
of the jet, and the corresponding energy in the fixed frame (frame of the central source)
is given by
\begin{equation}
\label{eq:en}
    \en\approx 0.1\,\Gamma\mbox{ GeV},
\end{equation}
where $\Gamma$ is the jet Lorentz factor.
As the emitted neutrinos propagate large distances to the observer at earth,
their energies are reduced by the cosmological redshift $(1+z)$. Pion decay produces
muon and electron neutrinos, however, due to flavor oscillations on the way to
the observer, neutrinos come in all of the three flavors.


\vspace{4 mm}
\noindent
{ \bf 3. IceCube+DeepCore capabilities for GRB detection} ---
%
While IceCube itself was mainly designed to observe
neutrinos with energies above 100\,GeV, it has been complemented with the component named DeepCore, a smaller Cherenkov detector with a higher concentration of optical modules, which targets neutrinos with energies down
to $\sim 10$~GeV   \cite{2011arXiv1109.3262A,2012APh....35..615A}.
 IceCube+DeepCore is most sensitive of all neutrino detectors (existing or planned)
in the energy range between 10 and 100~GeV  \cite{2011arXiv1109.3262A}.

Given a neutrino fluence $\Fn$ [cm$^{-2}$], the mean expectation for the number
of detected neutrinos is determined by the  detector effective area $\A$,
\begin{equation}
    \nD=\A \Fn.
\end{equation}
The effective area for IceCube+DeepCore was evaluated by \cite{2012APh....35..615A}. In the 10-100\,GeV energy range, their results can be approximately described by a power law,
\begin{equation}
   \A(\en)\approx 40 \left(\frac{\en}{100\rm ~GeV}\right)^2 {\rm ~cm^2}
\end{equation}
for muon neutrinos. (For electron neutrinos, the effective area is about two times smaller.)

The detector background (for up-going events) is dominated by atmospheric
neutrinos generated by cosmic rays from the northern hemisphere. We approximate the energy distribution of {\it detected} background neutrinos to be flat in the range of $10-100$\,GeV (e.g., \cite{2009arXiv0907.2263W}; c.f. \cite{2011JPhCS.335a2056F}). For $1.6\pi\,$sr region of the northern hemisphere we adopt a muon neutrino background rate of
\beq
   \frac{d\dot{n}^{\rm BG}}{d\en}\approx 100 {\rm ~GeV}^{-1}\,{\rm yr}^{-1}.
\eeq
The net background rate integrated over the spectral window below 100\,GeV
is $\dot{n}^{\rm BG}\approx 10^4$\,yr$^{-1}$.
Then the mean expectation for the background neutrino number in a single GRB is $\nBG\approx 10^{-2}(T/30\,\mbox{s})$,
where $T$ is the time interval during which most (e.g. 90\%) of the burst fluence comes; typically, $T\lesssim 10-20$~s (e.g., \cite{2012ApJS..199...18P}).
The background is small if the mean expectation for neutrino signal $\nD\gg \nBG$.

The background can be significantly reduced if we use the known location of the burst on the sky. GRBs are typically well localized by gamma-ray observations,
and many of the background neutrinos can be rejected using their directions. We adopt an uncertainty of $\sim5^{\circ}$ in the direction reconstruction of IceCube+DeepCore muon neutrinos, which is the nominal value at energies $\epsilon_\nu\sim100\,$GeV (\cite{2011ICRC....5..141D}; the uncertainty is somewhat greater at lower energies).
Assuming that the GRB is localized on the sky with a similar or better accuracy, the muon-neutrino background is effectively reduced by a factor of $\sim 1/200$:
\begin{equation}
\label{eq:BGmu}
  \nBGmu \approx 5\times10^{-5}  \left(\frac{T}{30 \rm ~s}\right).
\end{equation}
In our analysis, we adopt this average value for the entire energy range.

The direction reconstruction for electron neutrinos is difficult, which makes their
effective background level much higher. For this reason, we will focus below on the detection of muon neutrinos.

\vspace{4 mm}
\noindent
{\bf 4. Expected detection rate} --- 
For a burst with gamma-ray energy fluence $\S$\;[erg\,cm$^{-2}$] the expected number fluence of muon neutrinos is given by
\beq
    \Fmu=\frac{\eff \S}{3\en},
\eeq
where $\eff$ is the ratio of the burst energies emitted in neutrinos and gamma-rays
(\Eq~\ref{eq:w}), and $\en$ is the average energy of the GRB neutrino reaching the earth.
The factor of 1/3 takes into account that
the emitted neutrinos come mixed in three flavors, as a result of neutrino oscillations.
The mean expectation for the number of detected muon neutrinos is given by
\beq
\label{eq:nDmu}
   \nDmu=A\Fmu\approx 8\times 10^{-4} \eff \,
               \S_{-5}\,\left(\frac{\en}{100 \rm~GeV}\right)^2,
\eeq
where $S_{-5}=S/10^{-5}\,{\rm erg\,cm}^{-2}$. The observed neutrino energy is
reduced by the cosmological redhsift $(1+z)^{-1}$ from the value given
by \Eq~(\ref{eq:en}),
\beq
\label{eq:en1}
   \en\approx 30\,\left(\frac{\Gamma}{600}\right)\left(\frac{1+z}{2}\right)^{-1}  {\rm ~GeV}.
\eeq

Consider the example of a very bright burst GRB~080319B
\cite{2008Natur.455..183R}. Its gamma-ray fluence was
$\S\approx6.2\times10^{-4}$~erg~cm$^{-2}$ and its source was located at $z=0.937$, in the
northern hemisphere.
It had an isotropic-equivalent gamma-ray energy $\Eiso\sim10^{54}$\,erg.
The exact Lorentz factors of GRB jets are unknown, however it is expected that
the brightest bursts have particularly high $\Gamma\sim 10^3$
(which helps avoid gamma-gamma absorption and explain the observed GeV gamma-rays).
Then we find for GRB~080319B,
$\nDmu\approx 1.4\times 10^{-2} w\,\Gamma_3$, where $\Gamma_3=\Gamma/10^3$.

We conclude that the detection probability for an individual GRB is small unless
the burst occurs so close to us that its fluence has a huge value
$S>10^{-2}$\,erg\,cm$^{-2}$.

\begin{figure}
\resizebox{0.475\textwidth}{!}{\includegraphics{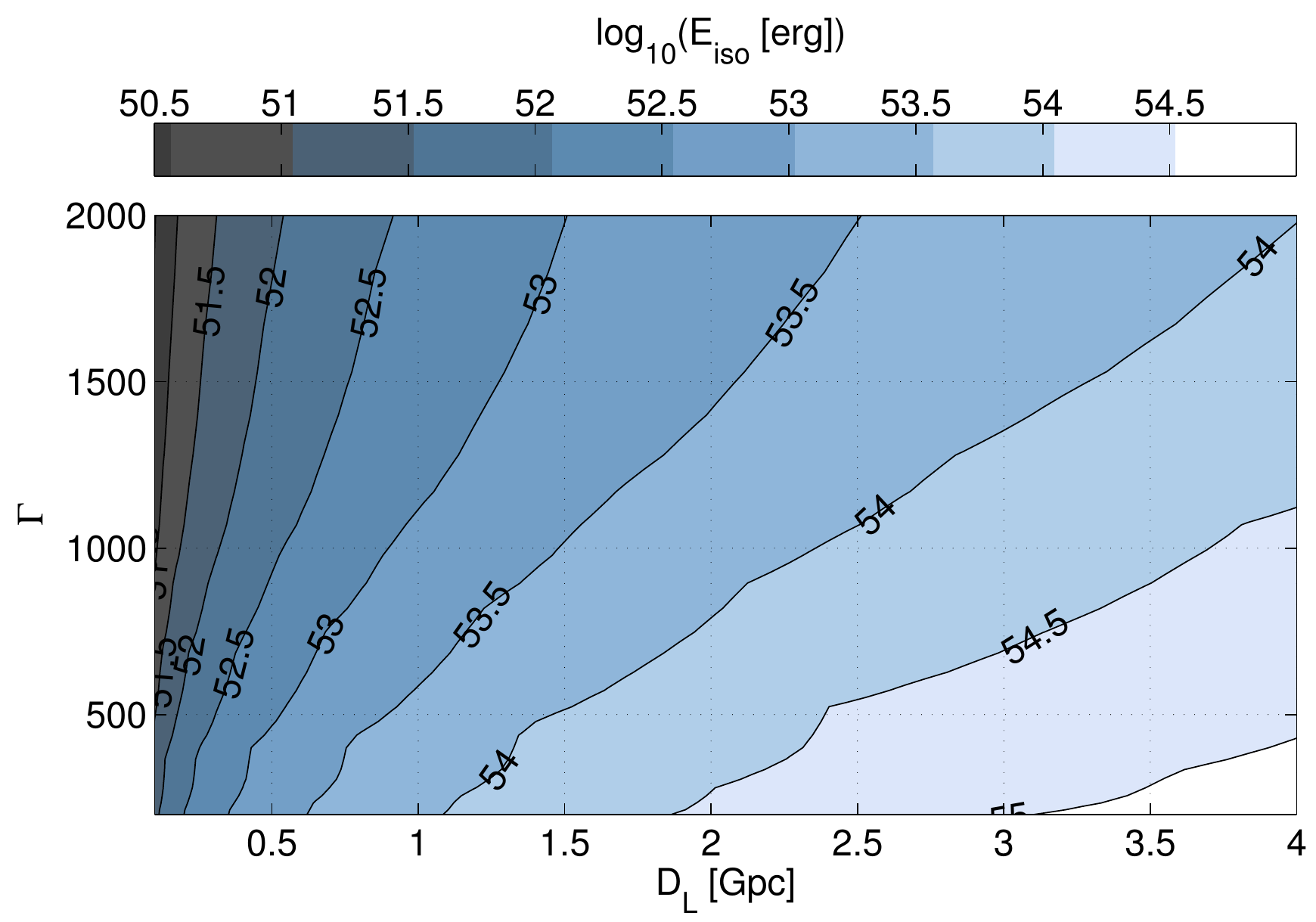}}
\caption{GRB isotropic-equivalent gamma-ray energy $\Eiso$ that would produce, on average, $1$ detected muon neutrino in IceCube+DeepCore, as a function of the GRB's
luminosity distance $D_L$ and Lorentz factor $\Gamma$. }\label{figure:singledetection}
\end{figure}

Figure~\ref{figure:singledetection} shows $\Eiso$
required to produce, on average, 1 detected neutrino in IceCube+DeepCore, as a
function of luminosity distance $D_L$ and Lorentz factor $\Gamma$.
One can see that the burst with a typical $\Eiso\sim 10^{53}$\,erg needs to be
within $\sim1$~Gpc to produce $\nDmu\gtrsim1$.

The mean expectation for detected neutrinos (\Eq~\ref{eq:nDmu}) is proportional to
the gamma-ray energy fluence $S$, which can be greatly increased if we consider a large sample of GRBs and add
their fluences together. Adding a burst to the sample is useful as long as it adds more
signal than background, i.e. if it contributes $\nDmu>\nBGmu$. This requires a minimum
fluence of the burst, which we find by comparing \Eqs~(\ref{eq:BGmu}) and (\ref{eq:nDmu}),
\beq
    \S_{\min}\approx \frac{7\times 10^{-6}}{\eff}\,\left(\frac{\en}{30 \rm~GeV}\right)^{-1}
                                 \left(\frac{T}{30 \rm ~s}\right)  {\rm erg~cm}^{-2}.
\eeq
The observed distribution of $S$ significantly flattens at $\S<10^{-5}$\,erg\,cm$^{-2}$, and adding these weaker bursts to the sample does not add much fluence. Thus, we can choose the sample by requiring
\beq
 \S>\Scut\sim 10^{-5} {\rm~erg~cm}^{-2},
\eeq
without losing much signal while still having a weak background $\nBGmu\ll\nDmu$.

\begin{figure}
\resizebox{0.475\textwidth}{!}{\includegraphics{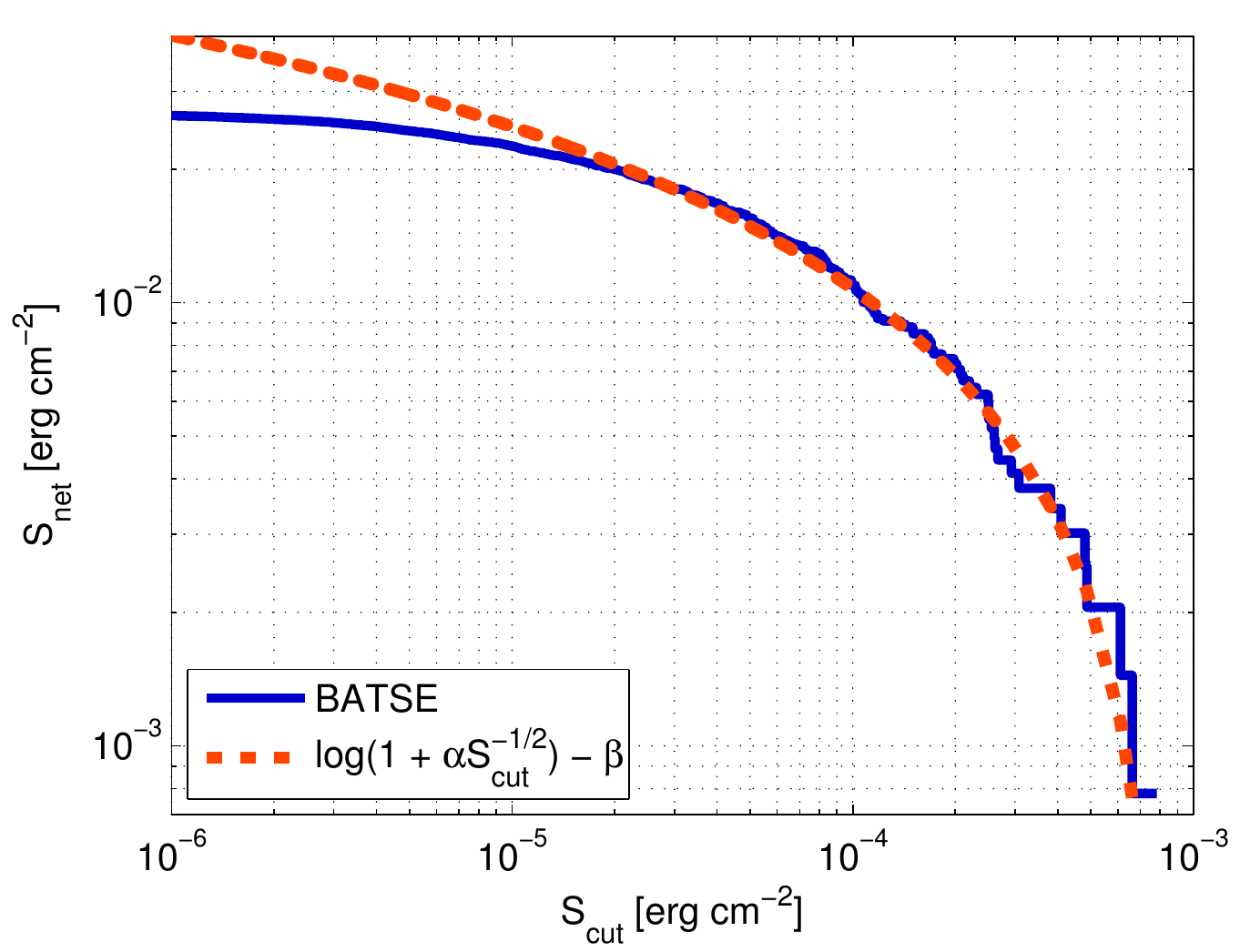}}
\caption{Net fluence $\Snet$ of GRBs with individual fluences $\S>\Scut$ as a function of $\Scut$ (solid line). Dashed line shows the analytical expression~(\ref{eq:Snet}).
}
\label{figure:FluenceDist}
\end{figure}

First, consider all bursts detected by the Burst and Transient Source Experiment
(BATSE; \cite{1992Natur.355..143M}) during its $\sim9$ years of operation.
The number of BATSE bursts with fluences $\S>\Scut\approx 10^{-5}$~erg~cm$^{-2}$ is
$N\approx 450$ \footnote{http://www.batse.msfc.nasa.gov/batse/grb/catalog/current/}, Figure~\ref{figure:FluenceDist} shows the net fluence $\Snet$ of bursts with individual fluences $\S>\Scut$, as a function of $\Scut$. For sufficiently high $\Scut$ of interest, the observed dependence of $\Snet$ on $\Scut$ may be approximated by
the following functional form,
\beq
 \label{eq:Snet}
   \Snet\propto \log(1+\alpha \Scut^{-1/2})-\beta,
\eeq
with $\alpha\approx0.057$ and $\beta\approx0.015$. We use $\Scut=10^{-5}$\,erg\,cm$^{-2}$, which gives $\Snet\approx 2.3\times 10^{-2}$\,erg\,cm$^{-2}$ (Figure~\ref{figure:FluenceDist}).

Half of the observed $\Snet$ comes from the northern hemisphere. Substituting $S=\Snet/2$ into \Eq~(\ref{eq:nDmu})
we find the mean expectation for the number of detected neutrinos for the BATSE sample,
\beq
   \nDmu\approx 1\,\overline{w\en_2}  \qquad {\rm (BATSE)},
\eeq
where $\en_2=\en/100$\,GeV and the line over $w\en_2$ signifies averaging over
 the sample; $ \overline{w\en_2}\sim 1$ is expected.

Next, consider the bursts observed by the Fermi Gamma-ray Burst Monitor (GBM; \cite{2009ApJ...702..791M}) and the Swift Burst Alert Telescope (BAT; \cite{2005SSRv..120..143B}) between June 1, 2010  and June 1, 2012. IceCube and DeepCore already operated during this period.
We include GRBs that were observed in the northern hemisphere. For each burst we use its measured fluence to calculate its contribution to $\nDmu$.
If a GRB has been detected with multiple observatories, we choose observations at higher energies, which give a better estimate for the total gamma-ray fluence $S$. (Swift BAT is sensitive to photon energies only up to $150\,$keV, therefore its observations typically underestimate $\S$.) In this estimate, we choose a fixed $\Gamma=600$ and $z=1$ and find $\nDmu\approx 0.13$ for the two-year sample.

The present all-sky rate of GRB detections is about 325 per year, when
bursts from Swift, Fermi, and the 9-spacecraft Interplanetary Network
are considered \cite{IPN}. Although the majority of the present missions have virtually no limitation to their lifetimes, funding considerations may
eventually force their demise over the next decade.  In the near
future, the French-Chinese SVOM mission, the Japanese ASTRO-H, and
ESA's BepiColombo will have either dedicated GRB detectors or gamma-ray
detectors with burst-detection capability.

\vspace{4 mm}
\noindent
{\bf 5. Conclusions} --- 
We conclude that there is a good chance for detecting 10-100\,GeV neutrinos in 5-10\,years of observations with IceCube/DeepCore. Given the low level of expected background, the detection of a few neutrinos would have significant implications for GRB physics. It would confirm dissipative nuclear collisions in the jet and would determine the parameter $\eff$ that measures the efficiency of neutrino emission relative to the gamma-ray efficiency (\Eq~\ref{eq:w}). The energy of detected neutrino $\en$, combined with a measured cosmological redshift of the burst, would give a direct estimate for the Lorentz factor of the jet, $\Gamma\approx 100 (1+z) (\en/10{\rm ~GeV})$, a key parameter of GRBs.

\vspace{4 mm}
The authors thank Kohta Murase, Philipp Oleynik, Valentin Pal'shin, Carsten Rott and Maggie Tse for their help. IB and SM are thankful for the generous support of Columbia University in the City of New York and the National Science Foundation under cooperative agreement PHY-0847182. AB was supported by NSF grant AST-1008334.

\bibliographystyle{h-physrev}

\end{document}